\documentclass{jetpl}
\usepackage{graphicx,amssymb}
\twocolumn

\lat

\title{Searching for family-number conserving neutral gauge bosons from
extra dimensions }

\rtitle{Searching for family-number conserving neutral bosons \dots
}

\sodtitle{Searching for family-number conserving neutral bosons from extra dimensions
}

\author{
J.M.~Fr\`ere$^+$,
M.V.~Libanov$^*$,
E.Ya.~Nugaev$^*$\thanks{e-mail: emin@ms2.inr.ac.ru},
S.V.~Troitsky$^*$
}

\rauthor{
J.M.~Fr\`ere,
M.V.~Libanov,
E.Ya.~Nugaev,
S.V.~Troitsky
}

\sodauthor{
Fr\`ere, Libanov, Nugaev, Troitsky
}

\address{$^+$  Service de Physique Th\'{e}orique, CP 225,
Universit\'{e} Libre de Bruxelles, B--1050, Brussels, Belgium;\\
$^*$~Institute for Nuclear Research of the Russian Academy of Sciences,
60th October Anniversary Prospect 7a, Moscow 117312 Russia\\
}

\abstract{
Previous studies have shown how the three generations of the Standard
Model fermions can arise from a single generation in
more than four dimensions, and
how off-diagonal neutral couplings arise for gauge-boson Kaluza-Klein
recurrences.
These couplings conserve family number in the leading approximation.
While an existing example, built on a spherical geometry,
suggests a high compactification scale, we conjecture that the overall
structure is generic, and work out possible signatures at colliders,
compatible with rare decays data.
}


\begin{document}

\maketitle

\section{Introduction.}
\label{sec:intro}
One reason to invoke more than four space-time dimensions
is to obtain elegant solutions for several
long-standing problems of particle physics \cite{LED}
(see Ref.~\cite{RuUFN} for a review). In particular, in the
frameworks of ``large extra dimensions'', models have
been suggested \cite{LT,FLT} and studied \cite{FLTnu} where three
generations of the Standard Model (SM) fermions appear as three zero modes
localized in the four-dimensional core of a defect with topological number
three. When both fermions and Higgs boson are localized on a brane, the
overlaps of their wave functions may result in a hierarchical pattern of
fermion masses and mixings \cite{Schmaltz}. This occurs naturally in the
models under discussion \cite{LT}. If the gauge fields are not localized
on a brane (localization of them is a complicated issue
\cite{gaugelocaliz}), then their Kaluza-Klein (KK) modes mediate
flavour-changing processes. For the case of the compactification of two
extra dimensions on a sphere \cite{FLNT}, and with one particular pattern
of charge assignments, the constraints from a flavour violation were
discussed in Ref.~\cite{FLNTflavour}. A distinctive feature of the models
of this class is the (approximate) conservation of the family number. This
note aims to discuss, without appealing to a particular model, the
phenomenology of flavour-violating KK bosons in this class of theories, to
be searched in future experiments.

\section{Distinctive features of "single-generation" extra-dimensional
models}

If our four-dimensional world is nothing but a core of a topological
defect in $(4+D)$ dimensions, then specific interactions of matter fields
with the defect may induce localization of massless modes of these fields
inside the core of the defect. Identification of the SM fields with these
(almost massless, compared to the scale of the defect) modes allows the
extra dimensions to be large but unobserved (see Ref.~\cite{RuUFN} for a
review and list of references). In particular, the index theorem
guarantees the existence of $N$ linearly independent
chiral zero modes of each fermion field in the bosonic background with
topological number $N$. This suggests to use $N=3$ to obtain three
generations of the SM fermions from a single one in extra dimensions.
Quite non-trivially, the linear independence of the three modes results in
their different behaviour at the origin, which may give rise to a
naturally hierarchical pattern of masses of the fermions of three
generations. We concentrate here on the most elaborated example of two
extra dimensions ($D=2$), though our qualitative results hold for a more
involved case of higher dimensions as well.

With $D=2$, the two extra dimensions can be parametrized in terms of one
radial, $r$, and one angular, $\phi$, variables.
The location of the defect corresponds to $r=0$; we suppose that the
compactification preserves rotational invariance and allow $\phi $ to
vary from $0$ to $2 \pi $. The defect itself has a structure of the $U(1)$
vortex. The definition of $r$ and its maximal value depend on the
compactification scheme. The three light four-dimensional families of
particles arising from a single family in six dimensions are characterized
then by different winding properties in $\phi $: three families enumerated
by $n=1,2,3$
have the following wave functions:
\[
   \psi_n\sim f_n(r) e^{i (3-n) \phi}.
\]

These wave functions correspond, in four dimensions, to the {\em gauge}
eigenstates of the SM fermions. To the first approximation, both the
theory and the background possess rotational invariance (shifts in $\phi $
supplemented by $U(1)$ transformations). The fermion mass matrix
originates from a $\phi$-independent scalar field,
and is thus perfectly diagonal, while the mass spectrum results from the
(in principle calculable) overlap of the wave functions of the scalar and
fermions, giving the usual hierarchy between families. At this level, the
mass and gauge eigenstates coincide, the family number corresponds to the
six-dimensional angular momentum and is thus exactly conserved (note that
this still does not forbid processes where both quark and lepton flavours
change oppositely, e.g. $K\to\mu \bar e$). Mixing between fermions of
different species,
leading to the desired Kobayashi-Maskawa (K-M) matrix, arises as a
suppressed, second-order effect controlled by an auxiliary scalar field
with winding number one, which generates transitions between adjoining
generations.

We will be concerned here with the gauge interactions.
The lowest mode of the gauge bosons in four dimensions left massless by
the vortex localization of the fermions, eventually acquires mass by the
Brout-Englert-Higgs formalism. The electrically neutral such bosons stay
as usual diagonal in their interactions with the fermionic mass
eigenstates. The charge universality is provided by the fact that the
lowest mode of a gauge boson is constant in transverse dimensions and
overlap integrals of the normalized fermionic wave functions with this
mode coincide with each other. This is not the case for the higher KK
modes of the vector particles; their profiles in $(r,\phi )$ are
determined by
\[
A_{lm}(r,\phi )=a_{lm}(r){\rm e}^{im\phi },
\]
where $l=1,2,\dots$ and $-l\le m \le l$.  Non-trivial profiles $a_{lm}$
cause different overlaps with fermions of different families while non-zero
windings result in transitions between generations.

Angular excitation of, for example, the first KK mode of $Z$-boson
behaves in six dimensions as
\[
    Z' \sim e^{\pm i\phi}.
\]
After integration in extra dimensions we obtain
an effective four-dimensional Lagrangian with complex vector field $Z'$,
which generates "horizontal" transitions between families
in which the generation number changes by one unit.

Such transitions are of course severely
limited by the high mass of the excitations, but also, in the first
approximation (neglecting of the K-M mixing), they do conserve
the family number. For instance,
the following processes are possible:
\[
\begin{array}{c}
s+\overline{d}\Rightarrow  Z'\Rightarrow  s+\overline{d},\\
s+\overline{d}\Rightarrow  Z'\Rightarrow \mu+\overline{e},\\
s+\overline{d}\Rightarrow  Z'\Rightarrow\tau+\overline{\mu}
\end{array}
\]
The first process in the first order in $Z'$ exchange thus conserves
  strangeness (and only small corrections linked to Cabibbo mixing would
  affect this), but the second, while conserving "family number", is a
  typical flavour-changing neutral current (FCNC) interaction, violating
  both strangeness and electron number.
While the last reaction is only possible in collisions, the study of rare
  $K_L$ decay puts strong limits on the mass and coupling constant of the
  $Z'$ \cite{FLNTflavour}
(similar relations hold for the photon and gluon angular
  excitations).

For the time being, we wish to retain these main characteristics of the
model: families are associated to some "winding number", conserved in
excited boson exchanges up to small Kobayashi-Maskawa corrections. The
detailed spectrum and strength of coupling of the gauge boson excitations
will depend on the exact geometrical implementation. A fully worked-out
example was presented in details in Ref.~\cite{FLNTflavour}, leading however
to a particularly high mass spectrum.

We conjecture that the same structure would remain intact in other
implementations. In Ref.~\cite{FLNTflavour} we supposed that the wave
functions of fermions and the first KK mode of the gauge boson
overlap strongly. Then the effective Lagrangian for the interaction
between fermions and flavour-changing bosons
contains the same coupling constant, as interaction with
the lowest KK-modes, i.e. the usual gauge bosons.
However, in particular models the profiles of fermionic
wave functions can be shifted, which means more freedom in couplings.
Let us denote the absolute value of the overlap integral
in extra dimensions between the wave functions $\psi _i,\psi_j$
of the fermions of generations $i,j$ and the wave function $\psi_{Z'}$
of the $Z$ excitation as
\[
\left|\int \psi_{Z'}\psi_i\psi_jd^2x\right|=\kappa _{ij}.
\]
Then
$(\bar e,\mu)$-interaction through $Z'$ is described by:
\[
\frac{g_{EW}\kappa _{12}}{2 \cos\theta_W} Z'_\mu
\left[
{1\over 2}\bar
e \gamma_\mu \gamma_5 \mu - \left({1\over 2}-2\sin^2\theta_W \right)\bar
e \gamma_\mu \gamma_5 \mu \right].
\]
The structure of this term coincides with the interaction of $\bar e,e$ and
$Z$ in SM with the strength $g=g_{EW}\kappa _{12}$. Interactions of other
leptons and quarks arise in a similar way.

The main restriction on the mass scale of
the model with
$\kappa _{ij}\simeq \delta _{i,i+1}$ arises from the limit on the branching
ratio for the process $K_L\to {\bar\mu} e$. Taking into account that
$\kappa _{12}$ can be different from $1$,
the strongest restriction from
the rare processes gives \cite{FLNTflavour}
\[
M_{Z'}\gtrsim \kappa _{12} \cdot 100 {\rm TeV}.
\]
In the simplest case when all $\kappa _{ij}\sim \kappa\cdot\delta _{i,i+1}
$,
\[
\kappa \lesssim \frac{M_{Z'}}{100 {\rm TeV}}.
\]
The decay width of the excited Z and photon results mainly from their decay
into fermions (with the possibility of model-dependent additional scalar
decay channels), and, by simple counting of modes, is estimated as
\[
    \Gamma (Z')= \kappa^{2} \cdot \frac{M_{Z'}}{M_{Z}}
    \cdot12.5 \cdot \Gamma _{Z\rightarrow \overline{\nu} \nu} \cong
    \kappa^{2} \cdot \frac{M_{Z'}}{M_{Z}} \cdot 1.8 {\rm GeV}.
\]

Similarly, the width of the first photon angular excitation is given by
\[
\Gamma(\gamma ')=\frac{16}{3}\kappa^2\sin^22\theta_W\cdot
\frac{M_{\gamma '}}{M_Z}\cdot\Gamma_{Z \to
\bar{\nu}\nu}\cong\kappa^2\cdot\frac{M_{\gamma '}}{M_Z}\cdot {\rm 1.3 GeV}.
\]
The first KK excitation of the gluon is wider
due to the larger coupling constant,
\[
\Gamma (G')\cong\kappa ^2\frac{M_{G'}}{M_Z}\cdot {\rm 7.2GeV}.
\]
A typical value of  $\Gamma$ is of order $10^{-3}{\rm
GeV}$ for $\kappa \simeq 10^{-2}$ and $M_{Z'}=1{\rm TeV}$.
In what follows we will assume that the masses of all the FCNC bosons
are equal:
\[
M_{Z'}=M_{\gamma '}=M_{g'}=M,
\]
as in the case of spherical model of Ref~\cite{FLNTflavour}.

\section{Collider searches}

The vector bosons discussed here can, in principle, be observed at
colliders due to the flavour-changing decay modes into
$(\mu e)$ and $(\tau \mu)$  pairs. The corresponding process is very
similar to the Drell-Yan pair production. A typical feature
of the latter is
the suppression of the cross section with
increasing of the resonance mass
at a fixed  center-of-mass energy.

The flavour-changing decays of this kind have a distinctive signature:
antimuon and electron (or their
antiparticles)
with equal and large transverse momenta in the final state.

\begin{figure}[htb]
\includegraphics[width=70mm,height=4.5cm]{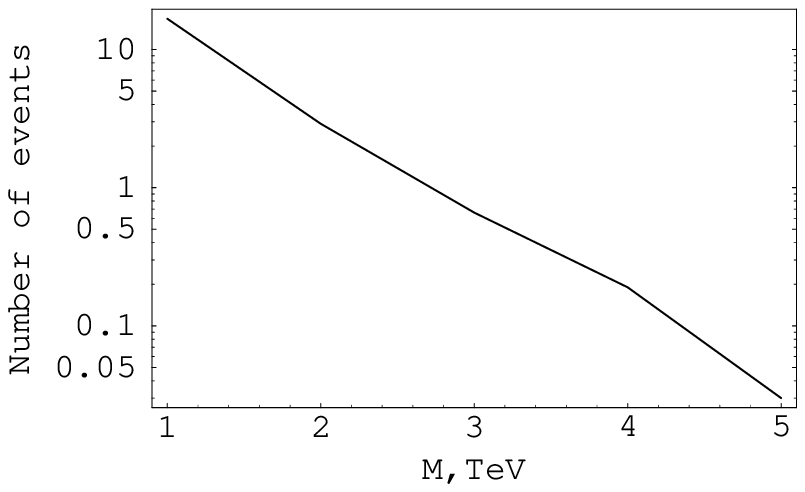}
\caption{{\rm Fig. 1}. Number of events for $(\mu^+e^-)$ pairs production
as a function of the vector bosons mass $M$, with  $\kappa=M/(100{\rm TeV})$.}
\label{f1}
\end{figure}

\begin{figure}[htb]
\includegraphics[width=70mm,height=4.5cm]{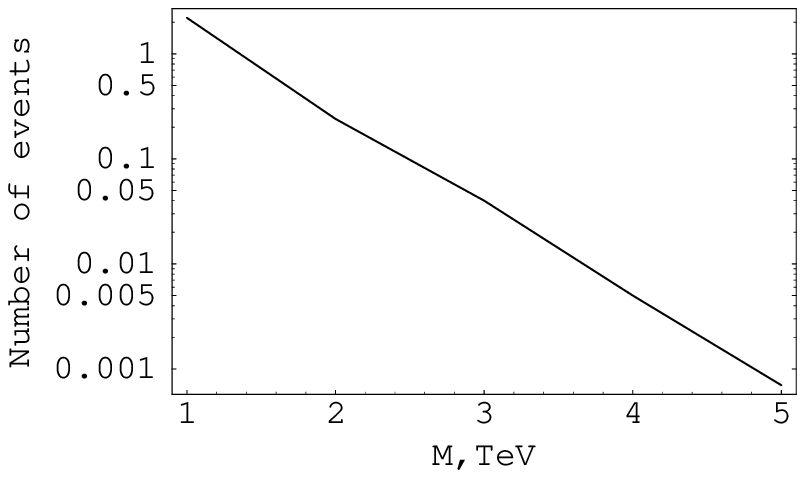}
\caption{{\rm Fig. 2}. Number of events for $(\mu^-e^+)$ pairs production
as a function of the vector bosons mass $M$, with  $\kappa=M/(100{\rm TeV})$.}
\label{f2}
\end{figure}

We estimate the number of events for the case of $pp$-collisions
with the help of the CompHEP package \cite{compHEP}. For our calculation we
use the expected LHC value of $100 {\rm fb}^{-1}$ for luminosity and
$\sqrt s =14{\rm TeV}$. The number of $(\mu^+e^-)$ events is presented
at Fig.\ref{f1} for different values of the vector bosons mass $M$
and $\kappa$ adjusted to  $\kappa=M/(100{\rm TeV})$. The same plot for
$(\mu^-e^+)$ pairs is given at Fig.\ref{f2}.

Note that
production of $(\mu^+e^-)$ pairs is more probable
than  $(\mu^-e^+)$
because the former process can use valence $u$ and $d$-quarks in the
proton, while the second only involves partons from the sea.
The same numbers are representative also for the $(\mu^-\tau^+)$ channel.

There are also other signatures of FCNC effects, in particular,
with hadronic final states, when $(\bar t,c)$ or $(\bar b, s)$ jets are
produced. The dominant contribution to these processes arises from the
interactions with higher KK modes of gluons, which have large coupling
constant. For the mass of $M_{G'}=1 {\rm TeV}$ we estimate the number of
events as $N=1.2\cdot 10^{3}$.
But potentially large SM backgrounds should be carefully considered for such channels.

\section{conclusions}
We have considered FCNC effects in models with approximate
family-number
conservation, mediated by the heavy
vector bosons in a class of models.
From our estimations, there is a reason for searching
for such FCNC bosons with masses of order $1 {\rm TeV}$ at LHC.
The main signature is the production of $(\mu^+e^-)$
or $(\mu^-\tau^+)$ pairs
with equal and large transverse momenta of leptons. Production of
$({\bar t}c)$ quarks is more probable, but less clear-cut due to the large background from SM processes.

On the other hand, the models with heavy vector bosons, whose interactions
conserve the family number, can be tested in experiments studing rare
processes. The strongest and the least model-independent limit on the mass
of these bosons arises from the limit on $K_L\to\mu^{\pm} e^{\mp}$
branching ratio (in this process, the family number does not change).
Discovery of this decay without signs of rare processes which violate the
generation number (such as $\mu e$-conversion) would support significantly
the models discussed here. Future experiments on the search of
lepton-flavour violating kaon decays are thus of great importance (see
Ref.~\cite{Landsberg} for relevant discussion).

We are indebted to S.~Demidov,  D.~Gorbunov, N.~Krasnikov,
L.~Landsberg and V.~Rubakov for numerous helpful
discussions. S.T.\ thanks Service de Physique Th\'{e}orique,
Universit\'{e} Libre de Bruxelles, where this work was
partially done, for warm
hospitality. This work is supported in part by the IISN (Belgium), the
``Communaut\'e Fran\c{c}aise de Belgique''(ARC), and the Belgium Federal
Government (IUAP); by RFFI grant 02-02-17398 (M.L., E.N.\ and S.T.); by
the Grants of the President of the Russian Federation NS-2184.2003.2
(M.L., E.N.\ and S.T.), MK-3507.2004.2 (M.L.) and MK-1084.2003.02 (S.T.);
by INTAS grant YSF~2001/2-129 (S.T.), by the grants of the Russian Science
Support Foundation (M.L and S.T.), by fellowships of the ``Dynasty''
foundation (awarded by the Scientific Council of ICFPM) (S.T. and E.N.)
and by the Project 35 of Young Scientists of the Russian Academy of
Sciences.

\end{document}